\documentclass%
[prb,%
superscriptaddress,%
showpacs,%
twocolumn]
{revtex4-1}

\usepackage{latexsym}        
\usepackage{amssymb}
\usepackage{amsmath}
\usepackage{amsfonts}
\usepackage{graphicx} 
\usepackage{textcomp}
\usepackage{longtable}

\begin{document}

\title{Relaxation dynamics induced in glasses by the absorption of hard X-ray photons}
							
\date{\today}

\author{G. Pintori}
\affiliation{Dipartimento di Fisica, Trento University, Povo, Trento I-38123, Italy}
\author{G. Baldi}
\affiliation{Dipartimento di Fisica, Trento University, Povo, Trento I-38123, Italy}
\author{B. Ruta}
\affiliation{ESRF--The European Synchrotron, CS40220, F-38043 Grenoble, France}
\affiliation{Univ. Lyon, Universit\'{e} Claude Bernard Lyon 1, CNRS, Lumi\`{e}re Mati\`{e}re, F-69622, Villeurbanne, France}
\author{G. Monaco}
\affiliation{Dipartimento di Fisica, Trento University, Povo, Trento I-38123, Italy}
\begin{abstract}
X-ray photon correlation is used to probe the slow dynamics of the glass-former B$_{2}$O$_{3}$ across the glass transition. In the undercooled liquid phase the decay times of the measured correlation functions are consistent with visible light scattering results and independent of the incoming flux;  in the glass they are instead temperature independent and show a definite dependence on the X-ray flux. This dependence can be exploited to obtain information on the volume occupied by the atoms that move in the glass following an absorption event. The length scale derived in this way, of the order of the nanometer, is consistent with that reported for the dynamical heterogeneities, suggesting the existence of a new scheme to get access to this fundamental quantity.
\end{abstract}

\maketitle

X-ray photon correlation spectroscopy (XPCS) is a powerful method for studying dynamics in disordered systems, giving access to length and time scales inaccessible by other techniques.
The typical time scales are much longer than those probed by inelastic X-rays scattering and the length scales much shorter than those investigated by visible-light techniques \cite{Grubel}.
XPCS requires the capability of third generation synchrotron radiation sources of producing coherent X-ray beams several orders of magnitude more intense than previously available \cite{Grubel}.
The pioneering papers by Sutton \textit{et al.} \cite{Sutton} and Brauer \textit{et al.} \cite{Brauer} laid the foundations of this research area.
XPCS has been used to study dynamics occurring on length scales $>$10 nm, for example, in small-angle scattering experiments on colloids and polymers, or near Bragg peaks of crystals \cite{Madsen}.
The capability to resolve single atomic motion in condensed matter with this technique has also been shown \cite{Stephenson, Leitner1}.

Much of the excitement about scattering with coherent X-rays, however, arises from the perspective to perform atomic resolution correlation experiments to study the complex dynamics of disordered
systems, whose archetypes are glasses. Their microscopic structure remains the object of active research.
Among the works that laid the groundwork in this field are those of Zachariasen \cite{Zach} and Warren \cite{Warren}.
The existence of short-range order in glasses was rather clear, while only the application of many different methods like neutron scattering, extended fine structure X-ray absorption and Raman spectroscopy \cite{Rao} has made it possible to provide hints into the medium-range order.
For what concerns the dynamics, the glassy state is described as arrested, with relaxation times too large to be observed on human time scales \cite{Ediger}. But what happens at the atomic scale?
The works by Ruta \textit{et al.} report the rather unexpected result that glasses display atomic rearrangements within few minutes in both metallic \cite{Ruta1} and silicate  glasses \cite{Ruta2},
and this even in the deep glassy state. While the dynamics of metallic glasses is intrinsic \cite{Giordano}, recent investigations clarify that this is not the case for silicon and germanium dioxide glasses where the atomic motion in the glassy state is induced by the X-ray beam \cite{Ruta3}.
Here we utilize XPCS to shed light on the effect of hard X-rays on the dynamics at the atomic level in the network glass B$_{2}$O$_{3}$ across the glass transition.
We show that this beam-induced dynamics competes with the structural relaxation, is negligible in the undercooled liquid phase and dominant in the glass.
The artificial dynamics induced by the beam can be described as a sequence of structural rearrangements involving the collective motion of up to thousands of atoms.

XPCS measurements on the B$_2$O$_3$ glass-former were performed at beamline ID10 at the ESRF in Grenoble (F), see Supplemental Material \cite{SM} for more details on the setup and on the sample preparation. The measurements were conducted by varying the flux of the incident beam, $F$, on the sample by means of different attenuators.
Each attenuator, made out of Si, leads to a decrease of the beam flux by a factor $\sim1/e$.
In particular, the atomic dynamics of the B$_{2}$O$_{3}$ glass was measured for: i) no attenuator, corresponding to an incoming beam flux $F_{0}= 8.6\cdot10^{10}$ ph/s per 200 mA current in the storage ring;
ii) a single attenuator filter, corresponding to a flux $F_{1} =2.6\cdot10^{10}$ ph/s per 200 mA; iii) and a double attenuator filter corresponding to a flux $F_{2}=9.8\cdot10^{9}$ ph/s per 200 mA.

\begin{figure}
\includegraphics[width=0.5\textwidth]{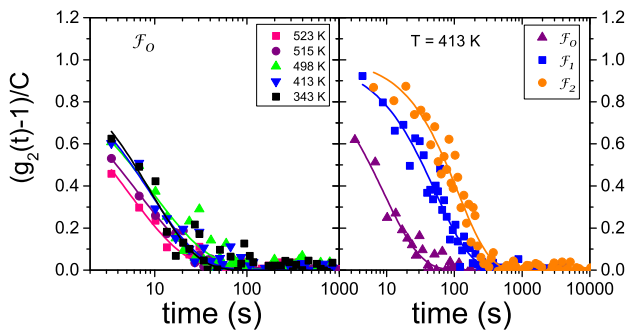}
\caption{\label{fig:func} (a) Normalized intensity auto-correlation functions (symbols) measured at Q$_{max}=1.5$ $\text{\AA}^{-1}$ in B$_{2}$O$_{3}$ for different temperatures across $T_g = 526$ K together with the best fitting stretched exponential line shapes. These measurements have been carried out with full beam flux, $F_{0}$. (b) Normalized intensity auto-correlation functions (symbols) measured at $T = 413$ K and Q$_{max}=1.5$ $\text{\AA}^{-1}$ for different incoming beam fluxes, see legend, together with the best fitting stretched exponential line shapes.}
\end{figure}

The intensity scattered by the B$_{2}$O$_{3}$ glass was collected for different temperatures in the 297--593 K range. At each temperature series up to 3,000 frames were taken with exposure times per frame,
$\Delta t_e$, in the range 2--7 s depending on the attenuator employed during the measurement. The recorded frames were subsequently analyzed by the multispeckle XPCS method \cite{Lumma, Chushkin} to obtain a set of temporal correlation functions. Fig. \ref{fig:func}(a) shows a series of normalized intensity autocorrelation functions, $g_{2}(Q,t)$,
measured in vitreous B$_{2}$O$_{3}$ by cooling the sample from the supercooled liquid phase to the glassy state ($T_{g}=526$ K) using the full beam flux $F_{0}$.
The dynamics becomes slower as the temperature is lowered down to 498 K, and shows very little temperature dependence at lower temperatures.
Fig. \ref{fig:func}(b) shows, moreover, that the atomic motion in the glassy state at $T = 413$ K strongly depends on the X-ray beam flux,
leading to an induced relaxation time that is shorter the higher is the incident beam flux, similar to what reported in Ref. \cite{Ruta3}. This effect is independent of the global dose released on the sample,
at least up to the maximum doses of $\simeq$ 2 GGy released on the same scattering volume during the measurements. This beam-induced effect is also not related to any visible structural damage: the scattered intensity, for instance, remains unaltered within about 2\% (see Supplemental Material \cite{SM}), and the beam-induced dynamics timescale reversibly changes with the incident flux, as also shown in Ref. \cite{Ruta3} for the case of SiO$_2$.

The shape of the correlation functions can be quantified by fitting to the data the Kohlrausch-Williams-Watts (KWW) expression \cite{Williams, Goetze}
\begin{eqnarray}
g_{2}\left(Q,t\right)=1+\mathcal{C}\left(Q\right)\cdot \exp \left[ -2\left(t\!/\tau\right)^{\beta} \right],
\label{eq:one}
\end{eqnarray}
where $\mathcal{C}=\mathcal{B}\left(Q \right) f_{Q}^{2}$ is the product of the experimental contrast and the square of the non-ergodicity factor; $\beta$ is the shape parameter and $\tau$ is the characteristic decay time.
Only few curves show the full decay from $1+\mathcal{C}$ to 1: most of them show in fact only the tail of the curve with a decay time that is fast on the scale fixed by $\Delta t_e$.
The fitting analysis of the experimental curves using Eq. (\ref{eq:one}) has then been carried out using all free fitting parameters ($\mathcal{C}$, $\tau$, $\beta$) only for the curves with longer $\tau$.
For these curves the parameter $\mathcal{C}$ comes out to be only little scattered around a mean value  $\mathcal{C}=\left(8.5\pm0.4\right) \times 10^{-3}$.
This value is lower than that observed in other glasses \cite{Ruta1,Ruta2}, because we had to use thicker samples in order to increase the scattered intensity (see Supplemental Material \cite{SM}).
Recalling that the non-ergodicity factor $f_{Q}$ in correspondence to the maximum of the structure factor is expected to display only a weak temperature dependence (e.g. see Ref. \cite{Ruta12}),
the fits to all experimental curves have been carried out using the previously mentioned fixed value for $\mathcal{C}$.

\begin{figure}
\includegraphics[width=0.5\textwidth]{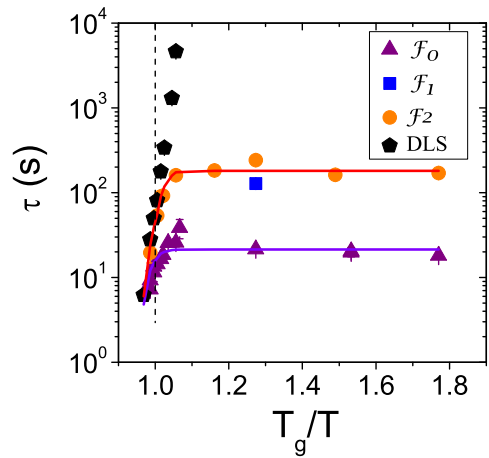}
\caption{Decay time measured using XPCS in B$_{2}$O$_{3}$ at Q$_{max}=1.5$ $\text{\AA}^{-1}$. Different symbols for the XPCS data refer to different X-ray beam fluxes, as reported in the legend. The black pentagons are macroscopic data obtained using dynamic light scattering \cite{Dallari}. The solid lines are obtained using Eq. (\ref{eq:model}) and refer to the XPCS data measured with the lowest and highest beam flux. The dashed vertical line marks the position of $T_g$.}
\label{fig:model}
\end{figure}

The temperature dependence of the decay time is reported in Fig. \ref{fig:model}. Macroscopic values (black squares) have been obtained from measurements carried out with dynamic light scattering \cite{Dallari}.
Different symbols for the XPCS data refer to different beam intensities, as reported in the legend. We highlight three main observations: i) Above $T_g$ the XPCS relaxation time, $\tau_{X}$, is very close to that measured in the visible range, and with very similar temperature dependence, confirming previous measurements on other systems~\cite{Ruta1,Ruta2,Evenson}.
ii) In the glassy state $\tau_{X}$ is almost temperature independent and remains in the 10--100 s range.
These findings are very similar to the behaviour recently observed in other network glasses \cite{Ruta2, Ruta3, Ross} at the atomic length--scale, contrary to the expectation of an almost arrested dynamics.
iii) While in the supercooled liquid region all of the XPCS data basically overlap, in the glassy state we observe different values of $\tau_{X}$ depending on the incident beam flux.
However, there are no clear evidences of radiation--damage (meaning permanent damage): the beam flux simply fixes the time--scale of the dynamics~\cite{Ruta3}.

In order to discriminate the beam-induced dynamics from the equilibrium dynamics, we can use a simple model where the decorrelation time measured with XPCS is written as:
\begin{eqnarray}
\frac{1}{\tau_{X}}=\frac{1}{\tau}+\frac{1}{\tau_{ind}}
\label{eq:model}
\end{eqnarray}
where $\tau$ is the structural relaxation time of B$_2$O$_3$, and $\tau_{ind}$ is the beam--induced decorrelation time.
We can use for $\tau_{ind}$ a temperature-independent (but beam-flux dependent) value given by $\tau_{X}$ in the glass;
and for $\tau$ the values obtained by photon correlation in the visible range and extrapolated below $T_g$. Fig. \ref{fig:model}
shows that this simple model (solid lines) describes very well the measured $\tau_{X}$ data. The beam induced dynamics takes place
in parallel to the spontaneous sample dynamics: above the glass transition temperature the structural relaxation is the fastest process and
therefore dominates, while below $T_{g}$ it becomes completely irrelevant.

\begin{figure}
\includegraphics[width=0.37\textwidth]{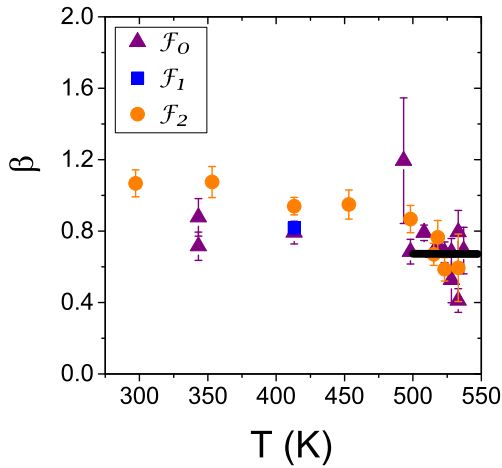}
\caption{Temperature dependence of the shape parameter $\beta$ for B$_{2}$O$_{3}$ at Q$_{max}=1.5$ $\text{\AA}^{-1}$. The different symbols refer to different beam fluxes, see legend. The full line indicates the mean value obtained with visible light scattering in the range 500--550 K \cite{Dallari}.}
\label{fig:par}
\end{figure}

The shape parameter, $\beta$, extracted from the KWW fits, is shown in Fig. \ref{fig:par}.
For the incoming beam flux $F_2$,
the obtained values for $\beta$ are basically temperature independent in the glass with a mean value of $\beta=0.97\pm0.04$.
At higher temperatures, they decrease to reach a value which is compatible with the average equilibrium value of $0.67\pm0.09$
obtained with visible light scattering \cite{Dallari} on a B$_2$O$_3$ sample prepared by exactly the same method as reported here. The reduction of $\beta$ in the XPCS data is therefore here another sign of the transition from a beam-induced dynamics in the glass to the equilibrium dynamics in the undercooled liquid.
The $\beta$ values at higher fluxes are affected by a considerable uncertainty because the decorrelation is faster and only a portion of the curve is measured.
Taking this into account, we conclude that the $\beta$ parameter does not show an appreciable dependence on the flux in the entire explored temperature range. It is however interesting to observe that, differently from the case of silica and germania \cite{Ruta3},
the shape parameter for the beam-induced decay corresponds to a simple exponential rather than a compressed ($\beta>1$) one.

\begin{figure}
\includegraphics[width=0.5\textwidth]{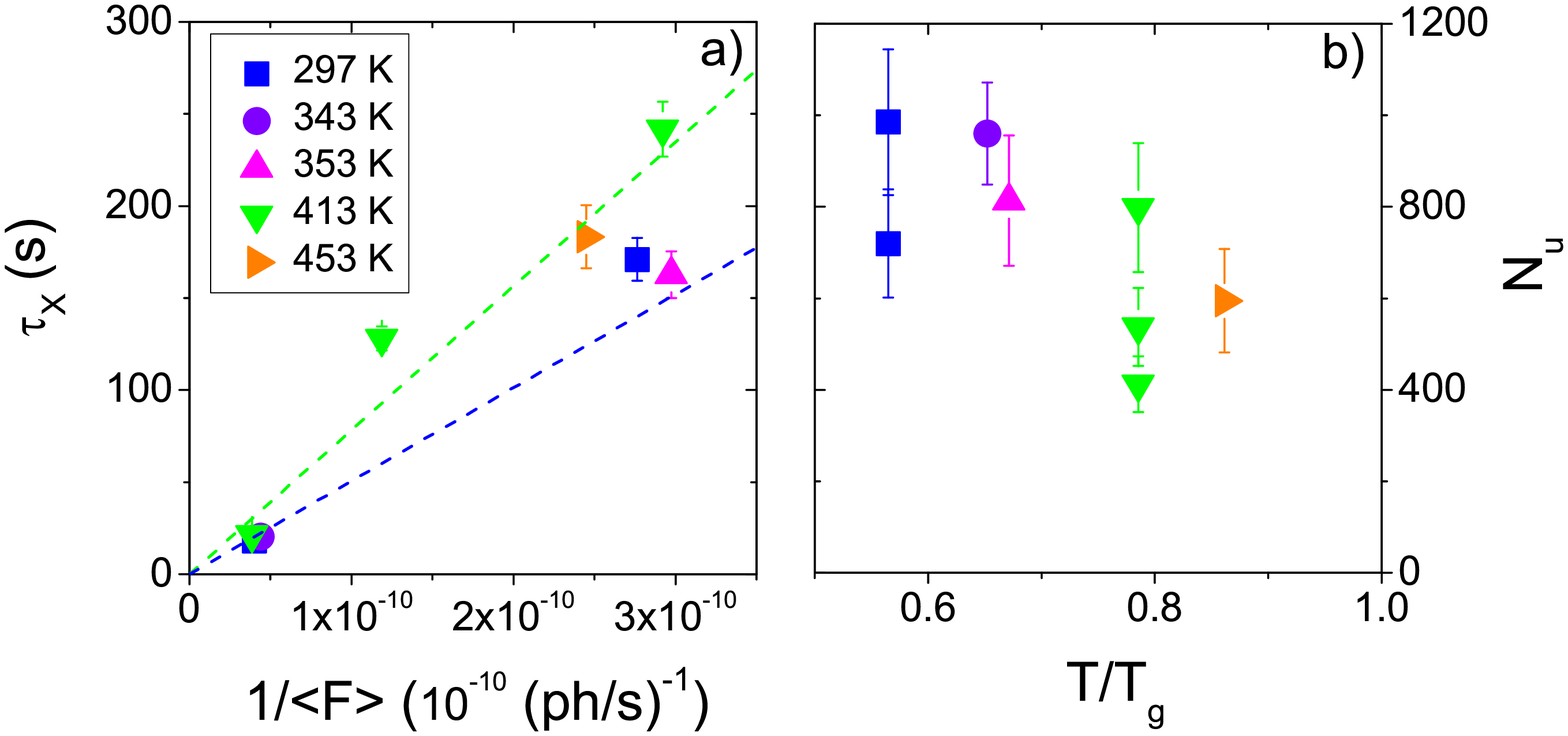}
\caption{(a) Decay time obtained from the XPCS measurements in B$_{2}$O$_{3}$ at Q$_{max}=1.5$ $\text{\AA}^{-1}$ as a function of the inverse of the average flux. Different symbols refer to different temperatures, see legend. The best fitting lines for $T=413$ K and $T =297$ K are also reported. (b) Temperature dependence of the number of B$_{2}$O$_{3}$ units that move following an X-ray absorption event, same symbols as in (a).}
\label{fig:flux}
\end{figure}

In the glassy state the decay time obtained by XPCS is only little temperature dependent and clearly decreases on increasing the X-ray beam flux, as shown in Fig. \ref{fig:model}.
In particular, our data are compatible with the expression $\tau_{X} \propto \langle F \rangle ^{-1}$, see the points corresponding to 413 K in Fig. \ref{fig:flux}(a). Here $\langle F \rangle = F\cdot \Delta t_e\!/\Delta t_l$ is the
average X-ray flux arriving at the sample; and $\Delta t_l$ is the lagtime, i.e. the sum of the exposure time $\Delta t_e$ and the readout time $\Delta t_r= 1.4$ s.
Fig. \ref{fig:flux}(a) confirms the results already reported in Ref. \cite{Ruta3} for the case of the silica and germania glasses.

We can also rephrase the previous observation by stating that the relaxation time measured by XPCS is inversely proportional to the average number of photons absorbed by the B$_2$O$_3$ sample, \textit{i.e.} $\tau \propto \langle F \rangle_a^{-1}$, where $\langle F \rangle_a = \langle F \rangle \left[1-\exp\left(-\mu L\right)\right]$, $\mu$ is the attenuation coefficient for B$_{2}$O$_{3}$ at 8.1 keV and $L$ is the sample thickness. It is then easy to clarify the meaning of this relation.
In fact, from the definition of intensity autocorrelation function, we know that in a time $\tau_X$, $N_{tot}\!/e$ of B$_2$O$_3$ units move by a distance $1/Q$, where $N_{tot}$ is the number of units in the scattering volume.
More precisely, in a time $\tau_X$, $f_Q N_{tot}\!/e $ of units move by a distance $1\!/Q$. However, $f_Q$ is very close to 1 when Q is close to the maximum of the structure factor (e.g. see Ref. \cite{Ruta12}), and therefore we can neglect its presence in what follows.
We also know that the number of photons absorbed in time $\tau_X$ is obviously $\langle F\rangle_a \cdot \tau_X$. Consequently, the number of units, $N_u$, that move after the absorption of one photon is the ratio of the number of units that move by a distance $1\!/Q$  in time $\tau_X$ and the number of photons absorbed in the same time:
\begin{eqnarray}
N_u=\frac{1}{e}\cdot\frac{N_{tot}}{\langle F\rangle_a \cdot \tau_X},
\label{eq:six}
\end{eqnarray}
where $N_u$ and $\tau_X$ can in principle be $Q$ dependent. The number $N_{tot}$ can be calculated using the sample mass density, $\rho$ = 1.83 g$\!/$cm$^{3}$, and the scattering volume defined by the beam spot size and the sample thickness.

The values for $N_u$ obtained in this way are reported in Fig. \ref{fig:flux}(b) as a function of temperature in the range where the observed dynamics is beam-induced,
i.e. for $T \leqslant 453$ K. It is interesting to remark that $N_u$ is large: $600 \pm 70$ B$_{2}$O$_{3}$ units, or $3000\pm200$ atoms. Eq. \ref{eq:six} is clearly a way to rationalize the flux dependence of the beam-induced decorrelation time measured in XPCS experiments: $N_u$ is the sample-dependent value that describes the proportionality of $\tau_X$ on the inverse average flux, and is the real outcome of XPCS measurements in beam induced conditions.
Note that the XPCS relaxation time depends on the scattering volume, being proportional to it. This simply reflects the fact that it takes longer to fluidize a larger amount of atoms.

It is interesting to explore the possibility that the $N_u$ units belong to the same volume $V_{c}$.
Assuming this volume being spherical, its radius $\xi$ will be related to $N_u$ by the relation $\xi = \sqrt[3]{3N_uv_{B_2O_3}\!/4\pi}$, where $v_{B_2O_3}$ corresponds to the volume of a B$_2$O$_3$ unit.
We obtain a value $\xi=2.3 \pm 0.1$ nm at $T=297$ K. It is suggestive to observe that this value is similar to those reported for the cooperativity length $\xi_{\alpha}$ at the
glass transition temperature ($\xi_{\alpha}$ = 2.0 nm \cite{Hong1} and 1.5 nm \cite{Hempel}).
We can hypothesise the following mechanism as responsible for the beam induced dynamics. The absorption of one photon generates a photoelectron which gives rise to a radiolysis-induced atomic displacement with a given probability~\cite{Griscom, Kinchin, Hobbs,Dapor}; however, since a glass is a metastable system characterized by internal stresses, this atomic displacement cannot be accommodated on its own and will be rather accompanied by the rearrangement of a larger region, corresponding to the cooperative volume.
A similar mechanism of stress release generated by random bond breaking has been recently exploited in numerical simulations of soft solids~\cite{DelGado} to probe their slow dynamics and was found responsible for the emergence of compressed correlation functions and of superdiffusivity.

It is also tempting to recognize, despite the large scattering of the present data, a temperature dependence for $N_u$, and thus for $\xi$, in Fig. 4b: $\xi$ is possibly decreasing on increasing $T$ as it is expected for the cooperative length measured in the liquid phase \cite{Adam, Donth}.
While this idea needs to be confirmed by experiments on more materials and validated in detail, it is  clear that alternative schemes can also be imagined to explain the value that we obtain for $N_u$. Considering in fact that the primary electrons produced by photoelectric absorption have an energy of about 8 keV and assuming a few tens of eV as the average energy loss per inelastic collision of the primary electron \cite{Egerton}, we can estimate that up to a few hundreds inelastic collisions per absorption event have the potential to give rise to atomic displacement by radiolysis. While this number is an order of magnitude smaller than the number of atomic displacements corresponding to $N_u$, it is possible (though unlikely) that all of these inelastic collisions give rise to the \AA-long displacements detected here, and therefore that each radiolysis event leads to the displacement of about ten atoms. Also in this scenario then there is some cooperativity required for the atomic displacements due to radiolysis, though clearly on a different length scale as the one discussed above.

In summary, we have investigated in some detail the effect of a hard X-ray beam on a borate glass using XPCS. In the supercooled liquid we probe the spontaneous dynamics related
to the structural relaxation at the atomic length-scale; in the glassy state, instead, the X-ray beam gives rise to a beam-induced dynamics. The X-ray beam thus fluidizes the sample: it induces local changes
and the overall configuration is renewed after the decay time $\tau_{X}$. These results confirm and extend to a new class of oxide glasses those already reported in \cite{Ruta3}; differently from that case, however, the shape of the correlation functions remains basically exponential instead than compressed. Moreover, the beam induced and the structural relaxation characteristic times are here shown to compete with each other and the two processes take place in parallel, so that the shorter one dominates the observed dynamics. We also confirm the proportionality between the induced-dynamics characteristic time and the inverse of the
average flux of the X-ray beam impinging on the sample reported in \cite{Ruta3}. We show here that this proportionality can be interpreted in terms of a fixed amount of material that rearranges after
one photon absorption event. The obtained value for this amount of material turns out to be similar to that expected for dynamical heterogeneities, and actually rather close to the available estimates for B$_2$O$_3$ \cite{Hong1, Hempel}. This observation,
when confirmed for other glasses, would establish a useful connection between the X-ray beam-induced dynamics here observed and a property of large interest for glasses.

The XPCS data here reported have been collected during one experiment at the ESRF (proposal HC1735).
We thank C. Armellini for help during the preparation of the sample. We acknowledge the ESRF for provision of synchrotron radiation facilities, and thank Y. Chushkin and K. L'Hoste for assistance in using beamline ID10.

\end{document}